\documentclass[english, authoryear]{elsarticle}
\usepackage{natbib}
\usepackage{url}
\usepackage[T1]{fontenc}
\usepackage[latin9]{inputenc}
\usepackage{babel}
\usepackage{array}
\usepackage{float}
\usepackage{multirow}
\usepackage{amstext}
\usepackage{amssymb}
\usepackage{graphicx}
\usepackage{rotfloat}
\usepackage{esint}
\PassOptionsToPackage{normalem}{ulem}
\usepackage{ulem}
\usepackage[unicode=true,
 bookmarks=false,
 breaklinks=false,pdfborder={0 0 1},backref=false,colorlinks=false]
 {hyperref}

\makeatletter

\providecommand{\tabularnewline}{\\}

\@ifundefined{date}{}{\date{}}
\usepackage{babel}
\journal{Journal}

\usepackage{lineno}
\usepackage{apalike}

\@ifundefined{showcaptionsetup}{}{%
 \PassOptionsToPackage{caption=false}{subfig}}
\usepackage{subfig}
\makeatother

\begin{document}
\begin{frontmatter}{}

\title{Financial Data Analysis Using Expert Bayesian Framework For Bankruptcy
Prediction}

\author[am]{Amir Mukeri}

\corref{cor1}

\ead{amir\_mukeri@acm.org}

\author[hs]{Habibullah Shaikh}

\author[dp]{Dr. D.P. Gaikwad}

\cortext[cor1]{Principal corresponding author}

\address[am]{Department of Computer Engineering, AISSMS College of Engineering,
No. 1, Kennedy Road, Pune, Maharashtra, India 411001. amir\_mukeri@acm.org }

\address[hs]{Credit Risk Consultant, D-2, Herms Grace, Camp, Pune, Maharashtra,
India 411001. habib80@gmail.com}

\address[dp]{Head of Department of Computer Engineering, AISSMS College of Engineering,
No. 1, Kennedy Road, Pune, Maharashtra, India 411001. dpgaikwad@aissmscoe.com }
\begin{abstract}
In recent years, bankruptcy forecasting has gained lot of attention
from researchers as well as practitioners in the field of financial
risk management. For bankruptcy prediction, various approaches proposed
in the past and currently in practice relies on accounting ratios
and using statistical modeling or machine learning methods. These
models have had varying degrees of successes. Models such as Linear
Discriminant Analysis or Artificial Neural Network employ discriminative
classification techniques. They lack explicit provision to include
prior expert knowledge. In this paper, we propose another route of
generative modeling using Expert Bayesian framework. The biggest advantage
of the proposed framework is an explicit inclusion of expert judgment
in the modeling process. Also the proposed methodology provides a
way to quantify uncertainty in prediction. As a result the model built
using Bayesian framework is highly flexible, interpretable and intuitive
in nature. The proposed approach is well suited for highly regulated
or safety critical applications such as in finance or in medical diagnosis.
In such cases accuracy in the prediction is not the only concern for
decision makers. Decision makers and other stakeholders are also interested
in uncertainty in the prediction as well as interpretability of the
model. We empirically demonstrate these benefits of proposed framework
on real world dataset using Stan, a probabilistic programming language.
We found that the proposed model is either comparable or superior
to the other existing methods. Also resulting model has much less
False Positive Rate compared to many existing state of the art methods.
The corresponding R code for the experiments is available at \href{https://github.com/amir1m/companies-bankruptcy-forecast}{Github repository}.
\end{abstract}
\end{frontmatter}{}

\section{Introduction}

The adverse effect of uncertainty across most of the global economies,
aggravated by the series of shocks to the financial sector, will likely
be felt increasingly across most sectors with ongoing pandemic as outlined by  Inetrnational Monetory Fund \Citep{long20206world}. Worsening conditions in the capital markets,
increasing economic instability and declining consumer confidence
has meant that sound liquidity positions, prudent treasury management
and resilient cash flows are important for credit quality, particularly
for speculative grade corporates.

Bankruptcy is a state wherein a corporation is unable to meet or repay
its statutory obligations from its operating activities. Financial
ratio analysis helps in identifying such types of bankrupt corporations
or corporations headed for bankruptcy in the near future.

Ratio Analysis is important for relevant stakeholders such as banks,
financial institutions, investors, and others in order to analyze
the financial position, liquidity, profitability, risk, solvency,
efficiency, and ultimately the health of any corporation. These are
quantitative metrics which help stakeholders in taking timely decisions
as well as identifying corporations which will go bankrupt. They are
early warning indicators of credit risk and financial distress. Broad
set of financial ratios would include liquidity ratios, solvency ratios,
profitability ratios, efficiency ratios, coverage ratios, and earnings
ratios. The numerator and denominator of the ratio to be calculated
are taken from the financial statements. Financial ratios can be seen
as a tool that can be used by every company to determine the financial
liquidity, the debt burden, and the profitability of the company and
whether it would be able to sustain and meet its obligations in the
near future. In fact, bond rating agencies like S\&P, Moody's and
Fitch also use financial ratio analysis to rate corporate bonds. These
ratios are critical component of any statistical or machine learning
model built to analyze and forecast financial distress.

There are various statistical techniques such as Linear Discriminant
Analysis (LDA), Multi-discriminant Analysis(MDA) as well as machine
learning techniques such as Support Vector Machines (SVM), Artificial
Neural Networks(ANN) used for bankruptcy prediction \Citep{devi2018survey}.
However none of these approaches allow quantification of uncertainty
in prediction or intuitive interpretability of model. These are specially
important in the fields such as healthcare diagnosis or finance where
the decision based on the predictions of models have critical implications.
On the other hand, Bayesian framework provides capability to include
expert judgement, uncertainty quantification as well as interpretability
of model right out of the box. The proposed model will help stakeholders
to be more vigilant, alleviate the above risks and in predicting bankruptcies.
The methodology presented in this article helps investors, banks,
financial institutions and other relevant stakeholders to take corrective
measures, timely decisions, strategic business planning in the funding
and investment process. In our knowledge this is the first of it's
kind study to apply Bayesian treatment to bankruptcy forecasting.

Reminder of the article is organized as follows. The second section
presents the literature survey covering some of the earlier work done
in the area of bankruptcy prediction, third section covers the dataset
used for conducting our experiments. Fourth section gives a brief
introduction to Generalized Linear Modeling method with underlying
mathematical underpinnings followed by sections on experimental results,
comparative performance with other methods and conclusions.

\section{Related Work}

Altman's Z-score \citep{altman1968financial} is a popular method
to determine whether a company would go bankrupt or not based on the
selected financial ratios. Vast literature on the study of distressed
companies relies on this method as first tool of choice available
in their toolbox. However it was not able to perform as expected as
demonstrated later in the comparative study section.

On the other side of spectrum, machine learning techniques such as
logistic regression was used by Ohlson \citep{ohlson1980financial}.
Qi Yu et al. \citep{yu2014bankruptcy} used Extreme Learning Machines
for the prediction which relies on single layer Artificial Neural
Network which is evolved with the knowledge of financial ratios to
discover the most optimum neural network structure. Arindam Chaudhary
et al. \citep{chaudhuri2017bankruptcy} proposed the soft computing
approach using fuzzy rough sets. Lahmiri, S et al. \citep{lahmiri2019can}
proposed the Generalized Regression Topology as an efficient method
to arrive at optimized neural architecture search. Barboza et al.
\citep{barboza2017machine} provides detailed survey on the statical
techniques such as Linear Discriminant Analysis (LDA) as well as machine
learning techniques such as Support Vector Machine and their comparative
analysis.

While each of the above methods have their merits and demerits none
of them provide uncertainty quantification of the prediction. On top
of that many models including those based on Artificial Neural Networks
(ANN) are black box models. They provide no opportunity to interpret
them and facilitate the discussion between experts, business stakeholders
even if they could perform well on test data. Secondly, many of these
models need the tuning of the hyperparameters, for example, number
of neurons or layers and learning rate which is a time consuming and
considered as black art \citep{snoek2012practical}.

Many of these challenges can be overcome with Bayesian modeling techniques.
Bayesian modeling techniques allows us to provide uncertainty quantification
in terms of posterior probability density \citep{gelman2013bayesian}.
While in the past there have been challenges in sampling from this
distribution due to computational limitations however with the increased
capacity and power of modern computation sampling has become much
more practical. Also advances in sampling algorithms such as Markov
Chain Monte Carlo (MCMC) have helped alleviate further the pain point
in this regard when exact posterior computation is intractable specially
for multivariate data.

\section{Dataset}

For this study, bankruptcy information of Polish companies \citep{Companie6:online}
is used which is based on dataset by Zeiba M et al. \citep{zikeba2016ensemble}
for the period between year 2000-2012. This dataset consists of various
accounting rations such as different types of assets to liabilities
in both long and short terms, and whether that company went bankrupt
or not. Such 10,000 sample data points are available. Dataset was
divided into training and test dataset with class label distribution
as shown in Table \ref{tab:Dataset}
\begin{center}
\begin{table}
\caption{\label{tab:Dataset}Class distribution in training and test dataset}

\centering{}%
\begin{tabular}{|c|c|c|}
\hline
Dataset  & \#Non-bankrupt  & \#Bankrupt\tabularnewline
\hline
\hline
Training  & 5487  & 113\tabularnewline
\hline
Test  & 2105  & 47\tabularnewline
\hline
\end{tabular}
\end{table}
\par\end{center}

All the 64 financial ratios available in the dataset are shown in
Appendix in the Table \ref{tab:Complete-list-of}

\section{Methodology}

Generalized Linear Model was chosen which is well known for its flexibility.
To implement Bayesian GLM regression model we have used R package\emph{
rstanarm} \citep{muth2018user} that provides an abstraction over
the underlying probabilistic programming language compiler of Stan.

Data is preprocessed by using standard scaling method. Then the variables
to be included in the model are selected along with suitable prior
distribution for model's priors. After building the models their predictive
performance was compared using K Fold cross validation. Once the model
is selected their significance is assessed using Bayesian hypothesis
testing methodology. in all the phases of these process domain expertise
was utilized.

This process is illustrated in the flowchart in Figure \ref{fig:Methodology}
\begin{center}
\begin{figure}
\includegraphics[scale=0.75]{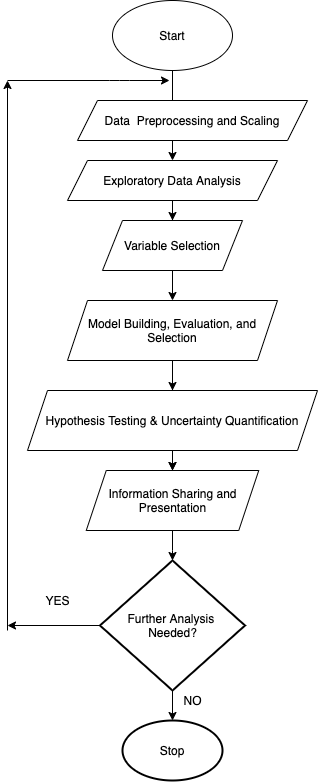} \centering{}\caption{\label{fig:Methodology}Methodology:In all the phases inputs from
domain expert are vital}
\end{figure}
\par\end{center}

\subsection{Generalized Linear Model}

Generalized Linear Models (GLM) are similar to linear regression models.
Here instead of having a normally distributed output variable it could
be made constrained, for example, for a binary logistic regression,
the output or response variable can be restricted to be between 0
and 1 or positive values only.

More formally, for binary outcome $y_{i}$, which is Bernoulli independent
and identically distributed (iid) given the probability of parameters
as $\theta_{i}$ ,

\begin{equation}
f(y_{i}|\theta_{i})\sim Bernualli(\theta_{i}),i=1,2,...n
\end{equation}
Bernoulli distribution is a special case of Binomial distribution
where out of $n$ inputs we model the probability of $k$ outputs
being 1 (success) or zero (failure).

The expected value for discrete random variable $X$ with a finite
number of outcomes $x_{1}\ensuremath{,}x_{2}\ensuremath{,...}x_{k}$
occurring with probabilities $\theta_{1}\ensuremath{,}\theta_{2}\ensuremath{,...}\theta_{k}$
respectively, is given by:
\begin{center}
\begin{equation}
\text{E[X]=\ensuremath{\sum_{i=1}^{k}x_{i}\theta_{i}}=\ensuremath{x_{1}\theta_{1}}+\ensuremath{x_{2}\theta_{2}}+...+\ensuremath{x_{k}\theta_{k}}}
\end{equation}
\par\end{center}

\begin{flushleft}
For a linear regression model, the expectation of $\theta_{i}$ is
given by:
\par\end{flushleft}

\begin{center}
\begin{equation}
E(\theta_{i})=\beta_{0}+\beta_{1}x_{i}
\end{equation}
\par\end{center}

\begin{flushleft}
However in order to constrain the value of $\theta_{i}$ to be between
0 and 1 we use \emph{log-odds ratio} as,
\par\end{flushleft}

\begin{center}
\begin{equation}
log(\theta)=log(\frac{\theta}{1-\theta})=\beta_{0}+\beta_{1}x_{i}
\end{equation}
\par\end{center}

This function, log of odds, is also called as logit-link function.
It called as\emph{ Link} since it helps to link given linear expression
to valid probability between 0 and 1 of successful outcome.

Therefore the expected value for the linear expression with probability
of success as $\theta_{i}$ and probability of failure as $(1-\theta_{i})$
is given as :
\begin{center}
\begin{equation}
E[\theta_{i}]=1*\theta_{i}+0*(1-\theta_{i})=\frac{e^{(\beta_{0}+\beta_{1}x)}}{1+e^{(\beta_{0}+\beta_{1}x)}}=\frac{1}{1+e^{-(\beta_{0}+\beta_{1}x)}}
\end{equation}
\par\end{center}

This model can be extended to multidimensional as well as multinomial
response variables by adding number of coefficients accordingly.

In order to build Bayesian linear regression model, we can select
priors on intercept and coefficient, for example, to be normally distributed
as,
\begin{center}
\begin{equation}
Intercept:\beta_{0}\sim Normal(a,b)
\end{equation}
\par\end{center}

\begin{center}
\begin{equation}
\beta_{1}\sim Normal(a,b)
\end{equation}
\par\end{center}

In case of \emph{Hierarchical Bayesian models} we can include the
priors for $a$ and $b$ as well.However in this study due to lack
obvious population subgroups in data such as sector or region hierarchical
model were not utilized.

In order to compute the posterior distribution of parameter vector
$\mathbb{\beta}$ using Bayesian inference,
\begin{center}
\begin{equation}
posterior=\frac{likelihood*prior}{evidence}
\end{equation}
\par\end{center}

\begin{center}
\begin{equation}
p(\beta|X)=\frac{f(X|\beta)*p(\beta)}{\int f(X|\beta)*p(\beta)d\beta}
\end{equation}
\par\end{center}

In above equation, denominator i.e. the normalizing constant is most
of the times a multidimensional integral and is analytically intractable
to compute. Therefore we approximate the posterior distribution using
Markov Chain Monte Carlo (MCMC) sampling algorithm. Probabilistic
programming language such as Stan \citep{carpenter2017stan} helps
in generating dependent samples from such distribution as an approximation
to the true posterior distribution in an efficient manner.

The posterior predictive distribution for a new example is given by,

\begin{equation}
p(y_{new}|X,\beta)=\int p(y_{new}|\beta)*p(\beta|X)d\beta
\end{equation}

\subsection{Variable Selection}

In the dataset there are total of 64 financial ratios that are explanatory
variables and a class variable indicating whether the company went
bankrupt or not indicated by 1 or 0 respectively. Expertise in credit
risk management available to us played a critical role in this phase.
We started with including variables in the model as shown in Table
\ref{tab:Variables-in-Model=00003D00003D00003D0000231} In Model\#1
with focus on ratios involving total liabilities.
\begin{center}
\begin{table}[H]
\begin{centering}
\caption{\label{tab:Variables-in-Model=00003D00003D00003D0000231}Ratios in
Model\#1}
\par\end{centering}
\begin{tabular}{|>{\centering}p{0.8\columnwidth}|}
\hline
Ratios\tabularnewline
\hline
\hline
{[}(cash + short-term securities + receivables - short-term liabilities)
/ (operating expenses - depreciation){]} {*} 365\tabularnewline
\hline
gross profit (in 3 years) / total assets\tabularnewline
\hline
(equity - share capital) / total assets\tabularnewline
\hline
(net profit + depreciation) / total liabilities\tabularnewline
\hline
operating expenses / total liabilities\tabularnewline
\hline
\end{tabular}
\end{table}
\par\end{center}

For the second model Model\#2 we included mostly short term liability
ratios as shown in Table \ref{tab:Variables-in-Model=00003D00003D00003D0000232}
\begin{center}
\begin{table}[H]
\begin{centering}
\caption{\label{tab:Variables-in-Model=00003D00003D00003D0000232}Ratios in
Model\#2}
\par\end{centering}
\begin{tabular}{|>{\centering}p{0.8\columnwidth}|}
\hline
Ratios\tabularnewline
\hline
\hline
book value of equity / total liabilities\tabularnewline
\hline
equity / total assets\tabularnewline
\hline
gross profit / short-term liabilities\tabularnewline
\hline
(inventory {*} 365) / sales\tabularnewline
\hline
operating expenses / short-term liabilities\tabularnewline
\hline
(current assets - inventory - receivables) / short-term liabilities\tabularnewline
\hline
profit on operating activities / sales\tabularnewline
\hline
(current assets - inventory) / short-term liabilities\tabularnewline
\hline
EBITDA (profit on operating activities - depreciation) / sales\tabularnewline
\hline
long-term liabilities / equity\tabularnewline
\hline
sales / short-term liabilities\tabularnewline
\hline
sales / fixed assets\tabularnewline
\hline
\end{tabular}
\end{table}
\par\end{center}

\subsection{Choice of Priors}

Since we believe that intercept as well coefficient should be close
to zero however there is chance that they could be large. Therefore
for both the models Student's t-distribution was used as it provides
fatter tails than a Normal distribution as prior for all parameters
with following hyperparameters:
\begin{center}
\begin{equation}
priors\sim Student\_t(df=7,location=0,scale=2.5)
\end{equation}
\par\end{center}

\subsection{Model Convergence}

Once the model is setup up, it is fitted on training data. In this
process we draw samples from posterior distribution using Markov Chain
Monte Carlo(MCMC) sampling, specifically using No U Turn Sampler (NUTS)
\citep{hoffman2014no}.

Both the models converged with 4000 total iterations with 2000 warmup
iterations with 4 chains. For all the coefficients we got an $\widehat{R}$
of less than 1.1.\emph{ }$\widehat{R}$ helps in analyzing the mixing
of chains, value of less than 1.1 is an indication of proper mixing
of chains \citep{vehtari2020rank} .

Analysis of posterior samples is shown in Table \ref{tab:Posterior-for-Model=00003D00003D00003D0000231}
for Model\#1 and in Table \ref{tab:Posterior-for-Model=00003D00003D00003D0000232}
for Model\#2.

\subsection{Model Selection}

Bayesian K Fold cross validation \citep{vehtari2017practical} is
used to compare both the models with 10 folds. The Expected Log Point
wise Density (elpd) difference between the two models is shown in
Table \ref{tab:ELPD-Difference}. The elpd helps in comparing the
model and hence in model selection. From the difference in the elpd
values we see that Model\#2, even though it has more number of variables
there is difference with high standard error. Therefore we choose
\emph{Model\#2} as better amongst the two models.

\begin{table}[h]
\begin{centering}
\caption{\label{tab:ELPD-Difference}ELPD Difference}
\par\end{centering}
\centering{}%
\begin{tabular}{|c|c|c|}
\hline
\multicolumn{1}{|c|}{} & ELPD difference  & Std Error\tabularnewline
\hline
\hline
Model\#2  & 0.0  & 0.0\tabularnewline
\hline
Model\#1  & -0.1  & 13.6\tabularnewline
\hline
\end{tabular}
\end{table}

Also Model\#1 as well Model\#2 was tested on unseen test dataset.Confusion
matrix from the predictions of Model\#1 and Model\#2 is shown in Table
\ref{tab:Model=00003D00003D00003D0000231Confusion-Matrix} and Table
\ref{tab:Model=00003D00003D00003D0000232Confusion-Matrix} respectively.
Model\#2 performs better than Model\#1 for the test data specifically
in detecting with positive cases. Here threshold of 0.5 for column
means of the posterior predictive samples was used to map the real
values between 0 and 1 to discrete 0s and 1s.

\begin{table}[h]
\caption{\label{tab:Model=00003D00003D00003D0000231Confusion-Matrix}Model\#1
Confusion Matrix}

\centering{}%
\begin{tabular}{|c|c|c|}
\hline
 & Predicted NO  & Predicted YES\tabularnewline
\hline
\hline
True NO  & 2101  & 4\tabularnewline
\hline
True YES  & 47  & 0\tabularnewline
\hline
\end{tabular}
\end{table}

\begin{table}[h]
\caption{\label{tab:Model=00003D00003D00003D0000232Confusion-Matrix}Model\#2
Confusion Matrix}

\centering{}%
\begin{tabular}{|c|c|c|}
\hline
 & Predicted NO  & Predicted YES\tabularnewline
\hline
\hline
True NO  & 2087  & 18\tabularnewline
\hline
True YES  & 39  & 8\tabularnewline
\hline
\end{tabular}
\end{table}

From the test results clearly Model\#2 performs better than Model\#1
on all measures.

\subsection{Bayesian Insight}

To determine if an attribute is significant or not we used ROPE (Region
of Practical Equivalence) \citep{kruschke2014doing}. ROPE value provides
the percentage of Credible Interval(CI) that is in the null region.
With sufficiently high value of ROPE the null hypothesis is accepted
otherwise rejected. The significant attributes are highlighted in
Table \ref{tab:Posterior-for-Model=00003D00003D00003D0000231}and
\ref{tab:Posterior-for-Model=00003D00003D00003D0000232}. Figure \ref{fig:Posterior-Density-plot}
shows posterior density plot using the drawn samples with uncertainty
intervals. We can see that the attributes with significant values
are away from zero.
\begin{center}
\begin{figure}
\includegraphics[scale=0.25]{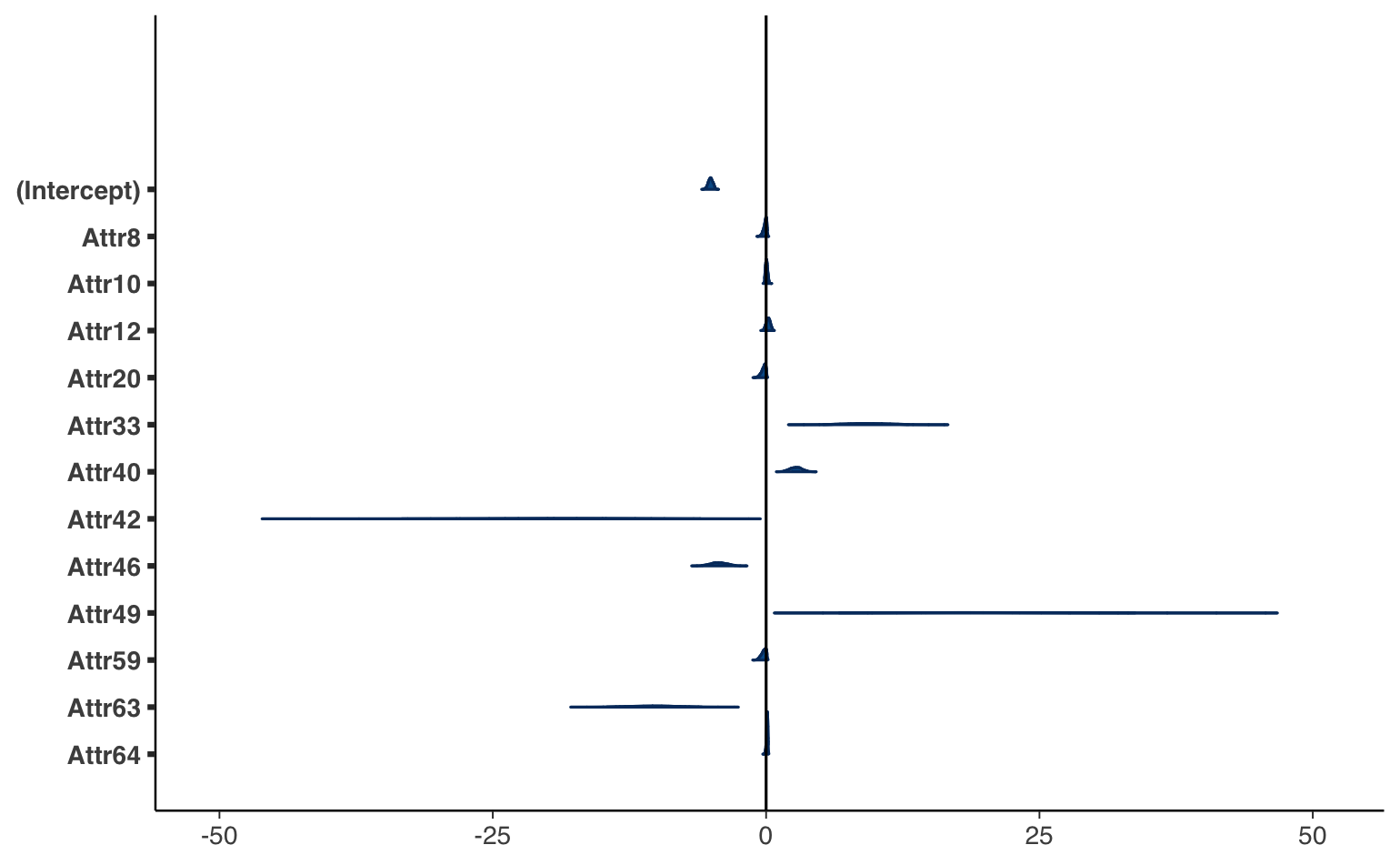} \centering{}\caption{\label{fig:Posterior-Density-plot}Posterior Density plot for Model\#1}
\end{figure}
\par\end{center}

From this analysis we can see that ratios involving short-term liabilities
are the best predictors for bankruptcy based on our dataset.

\begin{figure}
\includegraphics[scale=0.25]{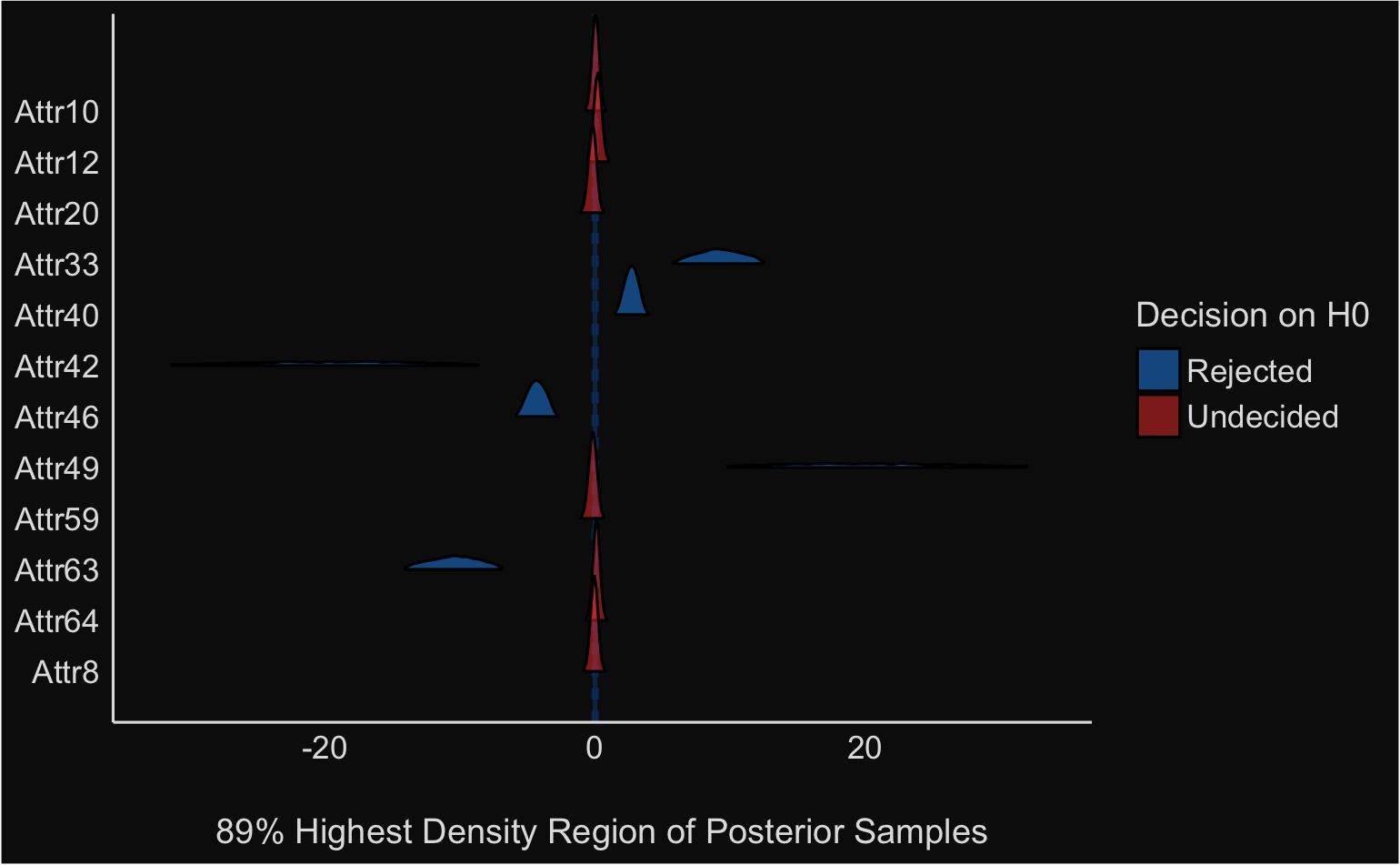}

\caption{\label{fig:Equivalence-Test-Plot}Equivalence Test Plot}
\end{figure}

We further analyze these ratios to get more insight using \emph{report
}\citep{makowski2019report} R package. After generating report we can examine
significance of an attribute. Following is the list of the analysis
of the attributes in Model\#2. Significant attributes in the below
summary are underlined.
\begin{itemize}
\item The effect of \uline{(Intercept)} has a probability of 100\% of
being negative and can be considered as large and significant (median
= -5.09, 89\% CI {[}-5.44, -4.72{]}, 0\% in ROPE, std. median = ).
The algorithm successfully converged (Rhat = 1.004) and the estimates
can be considered as stable (ESS = 1261).
\item The effect of Attr8 (book value of equity / total liabilities) has
a probability of 67.90\% of being negative and can be considered as
very small and not significant (median = -0.06, 89\% CI {[}-0.32,
0.12{]}, 76.15\% in ROPE, std. median = -0.06). The algorithm successfully
converged (Rhat = 1.001) and the estimates can be considered as stable
(ESS = 2197).
\item The effect of Attr10 (equity / total assets) has a probability of
63.68\% of being positive and can be considered as very small and
not significant (median = 0.04, 89\% CI {[}-0.12, 0.20{]}, 90.33\%
in ROPE, std. median = 0.04). The algorithm successfully converged
(Rhat = 1.000) and the estimates can be considered as stable (ESS
= 3208).
\item The effect of Attr12 (gross profit / short-term liabilities) has a
probability of 83.23\% of being positive and can be considered as
very small and not significant (median = 0.20, 89\% CI {[}-0.11, 0.50{]},
43.60\% in ROPE, std. median = 0.20). The algorithm successfully converged
(Rhat = 1.000) and the estimates can be considered as stable (ESS
= 3673).
\item The effect of Attr20 ((inventory {*} 365) / sales) has a probability
of 91.50\% of being negative and can be considered as very small and
not significant (median = -0.20, 89\% CI {[}-0.54, 0.05{]}, 45.88\%
in ROPE, std. median = -0.20). The algorithm successfully converged
(Rhat = 1.001) and the estimates can be considered as stable (ESS
= 3112).
\item The effect of \uline{Attr33 (operating expenses / short-term liabilities)}
has a probability of 100\% of being positive and can be considered
as large and significant (median = 9.22, 89\% CI {[}5.93, 12.31{]},
0\% in ROPE, std. median = 9.22). The algorithm successfully converged
(Rhat = 1.001) and the estimates can be considered as stable (ESS
= 2182).
\item The effect of \uline{Attr40 ((current assets - inventory - receivables)
/ short-term liabilities))} has a probability of 100\% of being positive
and can be considered as large and significant (median = 2.71, 89\%
CI {[}1.81, 3.54{]}, 0\% in ROPE, std. median = 2.71). The algorithm
successfully converged (Rhat = 1.001) and the estimates can be considered
as stable (ESS = 2086).
\item The effect of \uline{Attr42 (profit on operating activities / sales)}
has a probability of 100\% of being negative and can be considered
as large and significant (median = -19.73, 89\% CI {[}-31.69, -8.29{]},
0.03\% in ROPE, std. median = -19.73). The algorithm successfully
converged (Rhat = 1.000) and the estimates can be considered as stable
(ESS = 2313).
\item The effect of \uline{Attr46 ((current assets - inventory) / short-term
liabilities))} has a probability of 100\% of being negative and can
be considered as large and significant (median = -4.33, 89\% CI {[}-5.56,
-3.11{]}, 0\% in ROPE, std. median = -4.33). The algorithm successfully
converged (Rhat = 1.003) and the estimates can be considered as stable
(ESS = 1610).
\item The effect of \uline{Attr49 (EBITDA (profit on operating activities
- depreciation) / sales)} has a probability of 100\% of being positive
and can be considered as large and significant (median = 20.25, 89\%
CI {[}9.48, 32.78{]}, 0.03\% in ROPE, std. median = 20.25). The algorithm
successfully converged (Rhat = 1.000) and the estimates can be considered
as stable (ESS = 2360).
\item The effect of Attr59 (long-term liabilities / equity) has a probability
of 83.62\% of being negative and can be considered as very small and
not significant (median = -0.19, 89\% CI {[}-0.52, 0.10{]}, 49.45\%
in ROPE, std. median = -0.19). The algorithm successfully converged
(Rhat = 1.005) and the estimates can be considered as stable (ESS
= 1727).
\item The effect of \uline{Attr63 (sales / short-term liabilities)} has
a probability of 100\% of being negative and can be considered as
large and significant (median = -10.33, 89\% CI {[}-14.01, -7.09{]},
0\% in ROPE, std. median = -10.33). The algorithm successfully converged
(Rhat = 1.001) and the estimates can be considered as stable (ESS
= 2019).
\item The effect of Attr64 (sales / fixed assets) has a probability of 88.62\%
of being positive and can be considered as very small and not significant
(median = 0.10, 89\% CI {[}-8.84e-03, 0.21{]}, 93.80\% in ROPE, std.
median = 0.10). The algorithm successfully converged (Rhat = 1.002)
and the estimates can be considered as stable (ESS = 1703).
\end{itemize}
Figure \ref{fig:Equivalence-Test-Plot} shows the the plot of equivalence
test that helps in deciding weather the parameter should be accepted
or rejected based on 89\% Credible Interval (CI).

From above analysis, we conclude that declining sales and disproportionate
short-term liabilities are the main predictors of credit default in
near term future.

\section{Comparative Study}

Using Altman's Z-Score method \citep{altman1968financial} we found
that it was inadequate and gave a very low accuracy of 21\% on training
data and 20\% on test data with high False Positive Rate (FPR).

We also conducted experiments using other machine learning methods
such as Support Vector Machines, Random Forest, Neural Network, and
Logistic Regression.

SVM classifier with linear and radial basis kernels could not classify
a single true positive case from test data correctly. Random Forest
was able to classify one as true positive out of total 47 true positive
cases from test data.

\begin{figure}
\includegraphics[scale=0.35]{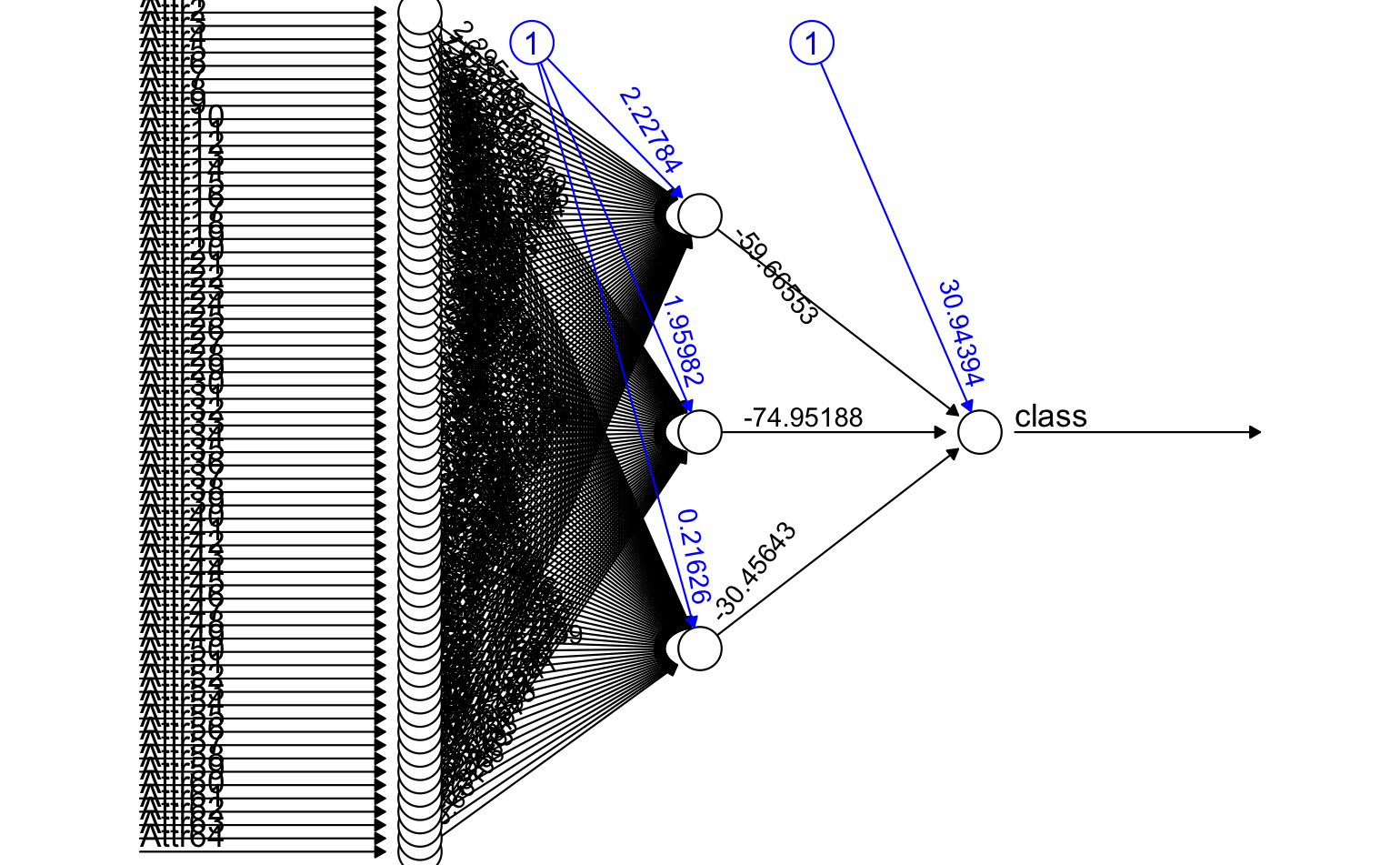}

\caption{\label{fig:Neural-Network}Neural Network}
\end{figure}

The experiments with neural network with 3 hidden layers were also
carried out. ANN is shown in Figure \ref{fig:Neural-Network} . As
shown in Table \ref{tab:Confusion-Matrix-NN} it was able to classify
8 out of 47 true bankrupt cases.

\begin{table}
\caption{\label{tab:Confusion-Matrix-NN}Neural Network Confusion Matrix}

\centering{}%
\begin{tabular}{|c|c|c|}
\hline
 & Predicted NO  & Predicted YES\tabularnewline
\hline
\hline
True NO  & 2087  & 18\tabularnewline
\hline
True YES  & 39  & 8\tabularnewline
\hline
\end{tabular}
\end{table}

\begin{table}
\caption{\label{tab:Confusion-Matrix-GLM}Non-Bayesian GLM Confusion Matrix}

\centering{}%
\begin{tabular}{|c|c|c|}
\hline
 & Predicted NO  & Predicted YES\tabularnewline
\hline
\hline
True NO  & 2088  & 17\tabularnewline
\hline
True YES  & 41  & 6\tabularnewline
\hline
\end{tabular}
\end{table}

\begin{figure}[H]
\subfloat[\label{fig:Accuracy}Accuracy]{\includegraphics[scale=0.45]{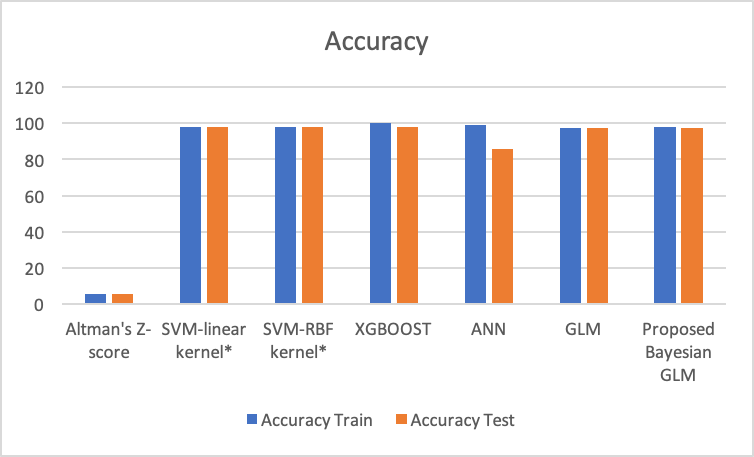}

}\subfloat[\label{fig:Precision}Precision]{\includegraphics[scale=0.45]{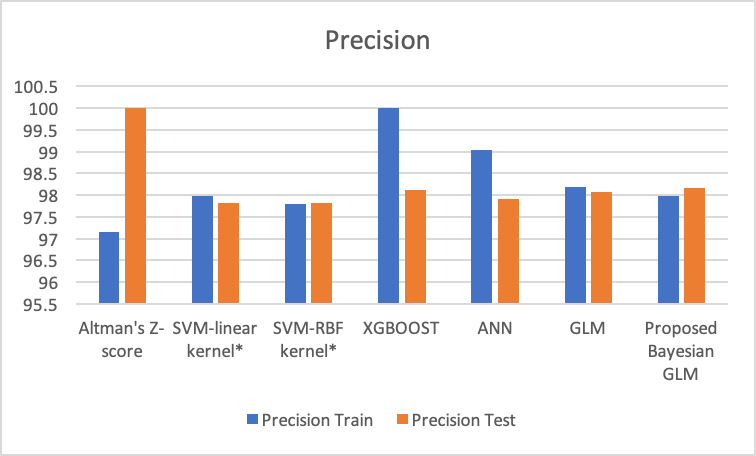}

}

\subfloat[\label{fig:Recall}Recall]{\includegraphics[scale=0.45]{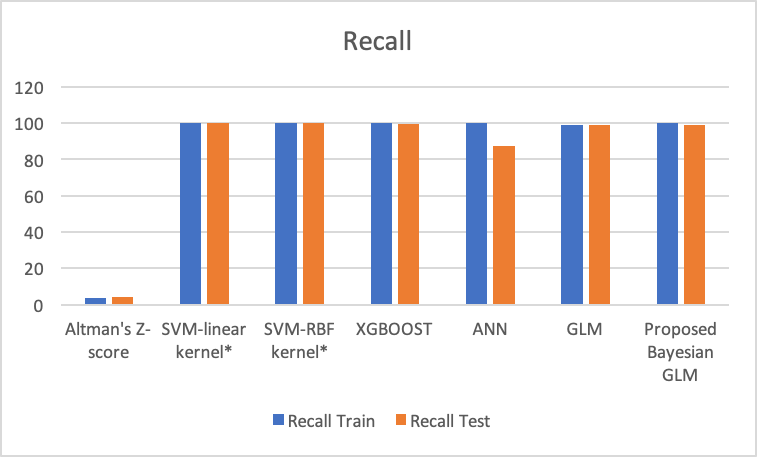}

}\subfloat[F1 Score]{\includegraphics[scale=0.45]{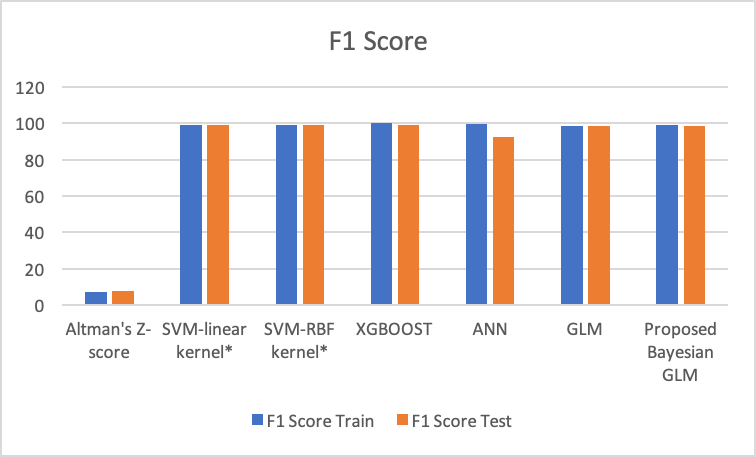}

}

\caption{\label{fig:Comparison-of-Performance}Performance Comparison}
\end{figure}

On the other hand, non-Bayesian Generalized Linear Model with logit
link function, we found the results with the test data as shown in
Table \ref{tab:Confusion-Matrix-GLM}. Given the flexibility of the
GLMs it was able to classify many more true positive cases correctly.
However it also ended up classifying more False positives in the process.XGBOOST(Extreme
Gradient Boosting Machine) was closest to the proposed Bayesian GLM
in detecting the positive cases. Performance metrics of various methodologies
that were tried against the proposed approach are listed in Table
\ref{tab:Comparison-Metrics}. Graphical visualization of the same
is shown in Figure \ref{fig:Comparison-of-Performance} Also it was
found that oversampling the minority class i.e. records with bankruptcy
class did not improve their performance any further for any of the
methods.

\begin{table}
\caption{\label{tab:Comparison-Metrics}Performance Comparison}

\begin{tabular}{|>{\raggedright}m{0.15\columnwidth}|>{\centering}p{0.06\columnwidth}|>{\centering}p{0.06\columnwidth}|>{\centering}p{0.06\columnwidth}|>{\centering}p{0.06\columnwidth}|>{\centering}p{0.06\columnwidth}|>{\centering}p{0.06\columnwidth}|>{\centering}p{0.06\columnwidth}|>{\centering}p{0.06\columnwidth}|}
\hline
\multirow{1}{0.15\columnwidth}{\emph{Methodology}} & \multicolumn{2}{c|}{\emph{Accuracy}} & \multicolumn{2}{c|}{\emph{Precision}} & \multicolumn{2}{c|}{\emph{Recall}} & \multicolumn{2}{c|}{\emph{F1 Score}}\tabularnewline
\hline
 & Train  & Test  & Train  & Test  & Train  & Test  & Train  & Test\tabularnewline
\hline
Altman's Z-score  & 5.6  & 6.0  & 97.16  & 100  & 3.75  & 4.0  & 7.22  & 7.7\tabularnewline
\hline
SVM-linear kernel  & 97.98  & 97.81  & 97.98  & 97.81  & 100  & 100  & 98.98  & 98.89\tabularnewline
\hline
SVM-RBF kernel  & 97.79  & 97.81  & 97.79  & 97.81  & 100  & 100  & 98.98  & 98.89\tabularnewline
\hline
XGBOOST  & 100  & 97.86  & 100  & 98.12  & 100  & 99.71  & 100  & 98.91\tabularnewline
\hline
ANN  & 99.05  & 86.01  & 99.04  & 97.92  & 100  & 87.55  & 99.51  & 92.45\tabularnewline
\hline
GLM  & 97.44  & 97.3  & 98.19  & 98.07  & 99.21  & 99.19  & 98.7  & 98.63\tabularnewline
\hline
Proposed Bayesian GLM  & 97.98  & 97.25  & 97.98  & 98.16  & 100  & 99.0  & 98.98  & 98.60\tabularnewline
\hline
\end{tabular}
\end{table}

Critics of Bayesian methodology have objected to inclusion of prior
knowledge as being subjective however other method viz. frequentist
do also use expert inputs without explicitly stating it \citep{brownstein2019role}.
Frequnetist models have no provision to explicitly include the same
while building a model. Bayesian framework allows the inclusion of
prior judgment explicitly and updating it based on the evidence as
we proceed in our experiments. This helps in directly driving decision
making process using data as well as taking benefit of an expert judgement.

\section{Conclusion}

With turbulent economic times ahead assessing and forecasting the
financial well being of commercial entities will gain more and more
traction. Most of the conventional approaches including Artificial
Neural Network are black box in nature and provide no help in interpretation
of model which is crucial in decision making process. The proposed
Bayesian GLM model and methodology meets the critical requirements
of interpretability as well as inclusion of expert judgement in all
the phases of model building process. Secondly, the proposed method
do not need any hyperparameter tuning or neural architecture search.
In this research, we built and tested model with this approach on
the bankruptcy data of Polish companies. It was found that specifically
looping in expert reduced the time for the model building and model
selection considerably compared to other combinatorial and computationally
intensive methods. Expert in the loop approach also helped in interpreting
the model and in communication of results of analysis to the relevant
stakeholders. At the end of the analysis we found that the ratios
involving short term liabilities were strong predictors of companies
going bankrupt. We believe this study will help open new avenues for
further exploration of novel applications of Bayesian methodology
in the areas of credit risk management and in investment decision
making along with inclusion of domain expertise in the process.

\appendix

\section{Exploratory Variables}
\begin{center}
\begin{table}[H]
\caption{\label{tab:Complete-list-of}Complete list of variables}

\begin{tabular}{|>{\centering}p{0.6\columnwidth}|>{\centering}p{0.6\columnwidth}|}
\hline
\centering{}Variable  & Variable\tabularnewline
\hline
\hline
attr1 - net profit / total assets  & attr33 - operating expenses / short-term liabilities\tabularnewline
\hline
attr2 - total liabilities / total assets  & attr34 - operating expenses / total liabilities\tabularnewline
\hline
attr3 - working capital / total assets  & attr35 - profit on sales / total assets\tabularnewline
\hline
attr4 - current assets / short-term liabilities  & attr36 - total sales / total assets\tabularnewline
\hline
attr5 - {[}(cash + short-term securities + receivables - short-term
liabilities) / (operating expenses - depreciation){]} {*} 365  & attr37 - (current assets - inventories) / long-term liabilities\tabularnewline
\hline
attr6 - retained earnings / total assets  & attr38 - constant capital / total assets\tabularnewline
\hline
attr7 - EBIT / total assets  & attr39 - profit on sales / sales\tabularnewline
\hline
attr8 - book value of equity / total liabilities  & attr40 - (current assets - inventory - receivables) / short-term liabilities\tabularnewline
\hline
attr9 - sales / total assets  & attr41 - total liabilities / ((profit on operating activities + depreciation)
{*} (12/365))\tabularnewline
\hline
attr10 - equity / total assets  & attr42 - profit on operating activities / sales\tabularnewline
\hline
attr11 - (gross profit + extraordinary items + financial expenses)
/ total assets  & attr43 - rotation receivables + inventory turnover in days\tabularnewline
\hline
attr12 - gross profit / short-term liabilities  & attr44 - (receivables {*} 365) / sales\tabularnewline
\hline
attr13 - (gross profit + depreciation) / sales  & attr45 - net profit / inventory\tabularnewline
\hline
attr14 - (gross profit + interest) / total assets  & attr46 - (current assets - inventory) / short-term liabilities\tabularnewline
\hline
attr15 - (total liabilities {*} 365) / (gross profit + depreciation)  & attr47 - (inventory {*} 365) / cost of products sold\tabularnewline
\hline
attr16 - (gross profit + depreciation) / total liabilities  & attr48 - EBITDA (profit on operating activities - depreciation) /
total assets\tabularnewline
\hline
attr17 - total assets / total liabilities  & attr49 - EBITDA (profit on operating activities - depreciation) /
sales\tabularnewline
\hline
attr18 - gross profit / total assets  & attr50 - current assets / total liabilities\tabularnewline
\hline
attr19 - gross profit / sales  & attr51 - short-term liabilities / total assets\tabularnewline
\hline
attr20 - (inventory {*} 365) / sales  & attr52 - (short-term liabilities {*} 365) / cost of products sold)\tabularnewline
\hline
attr21 - sales (n) / sales (n-1)  & attr53 - equity / fixed assets\tabularnewline
\hline
attr22 - profit on operating activities / total assets  & attr54 - constant capital / fixed assets\tabularnewline
\hline
attr23 - net profit / sales  & attr55 - working capital\tabularnewline
\hline
attr24 - gross profit (in 3 years) / total assets  & attr56 - (sales - cost of products sold) / sales\tabularnewline
\hline
attr25 - (equity - share capital) / total assets  & attr57 - (current assets - inventory - short-term liabilities) / (sales
- gross profit - depreciation)\tabularnewline
\hline
attr26 - (net profit + depreciation) / total liabilities  & attr58 - total costs /total sales\tabularnewline
\hline
attr27 - profit on operating activities / financial expenses  & attr59 - long-term liabilities / equity\tabularnewline
\hline
attr28 - working capital / fixed assets  & attr60 - sales / inventory\tabularnewline
\hline
attr29 - logarithm of total assets  & attr61 - sales / receivables\tabularnewline
\hline
attr30 - (total liabilities - cash) / sales  & attr62 - (short-term liabilities {*}365) / sales\tabularnewline
\hline
attr31 - (gross profit + interest) / sales  & attr63 - sales / short-term liabilities\tabularnewline
\hline
attr32 - (current liabilities {*} 365) / cost of products sold  & attr64 - sales / fixed assets\tabularnewline
\hline
\end{tabular}
\end{table}
\par\end{center}

\section{Posterior Description}
\begin{center}
\begin{sidewaystable}[H]
\begin{centering}
\caption{\label{tab:Posterior-for-Model=00003D00003D00003D0000231}Posterior
for Model\#1}
\par\end{centering}
\centering{}%
\begin{tabular}{|>{\centering}p{0.2\columnwidth}|>{\centering}p{0.08\columnwidth}|>{\centering}p{0.12\columnwidth}|c|>{\centering}p{0.15\columnwidth}|>{\centering}p{0.08\columnwidth}|c|>{\centering}p{0.08\columnwidth}|}
\hline
\centering{}Parameter  & Median  & 89\% CI  & pd  & 89\% ROPE  & \% in ROPE  & Rhat  & ESS\tabularnewline
\hline
\hline
(Intercept)  & -4.670  & {[}-4.951, -4.390{]}  & 1.000  & {[}-0.181, 0.181{]}  & 0.000  & 1.001  & 1150.992\tabularnewline
\hline
gross profit (in 3 years) / total assets  & 0.150  & {[}-0.033, 0.311{]}  & 0.915  & {[}-0.181, 0.181{]}  & 64.645  & 1.002  & 1779.815\tabularnewline
\hline
(equity - share capital) / total assets  & -0.160  & {[}-0.263, -0.043{]}  & 0.989  & {[}-0.181, 0.181{]}  & 65.319  & 0.999  & 6421.138\tabularnewline
\hline
(net profit + depreciation) / total liabilities  & -0.961  & {[}-1.262, -0.688{]}  & 1.000  & {[}-0.181, 0.181{]}  & 0.000  & 1.000  & 2235.707\tabularnewline
\hline
operating expenses / total liabilities  & 0.661  &  & 1.000  & {[}-0.181, 0.181{]}  & 0.000  & 1.000  & 1847.540\tabularnewline
\hline
{[}(cash + short-term securities + receivables - short-term liabilities)
/ (operating expenses - depreciation){]} {*} 365  & 0.380  & {[} 0.511, 0.802{]}  & 0.950  & {[}-0.181, 0.181{]}  & 29.739  & 1.002  & 1353.509\tabularnewline
\hline
(current assets - inventory) / short-term liabilities  & -2.163  & {[}-0.038, 0.897{]}  & 1.000  & {[}-0.181, 0.181{]}  & 0.000  & 1.001  & 1239.331\tabularnewline
\hline
\end{tabular}
\end{sidewaystable}
\par\end{center}

\begin{center}
\begin{table}[H]
\begin{centering}
\caption{\label{tab:Posterior-for-Model=00003D00003D00003D0000232}Posterior
for Model\#2}
\par\end{centering}
\centering{}%
\begin{tabular}{|>{\centering}p{0.25\columnwidth}|>{\centering}p{0.08\columnwidth}|>{\centering}p{0.12\columnwidth}|c|>{\centering}p{0.1\columnwidth}|>{\centering}p{0.07\columnwidth}|c|>{\centering}p{0.1\columnwidth}|}
\hline
\centering{}Parameter  & Median  & 89\% CI  & pd  & 89\% ROPE  & \% in ROPE  & Rhat  & ESS\tabularnewline
\hline
\hline
(Intercept)  & -5.078  & {[} -5.408, -4.727{]}  & 1.000  & {[}-0.181, 0.181{]}  & 0.000  & 1.002  & 1273.642\tabularnewline
\hline
book value of equity / total liabilities  & -0.064  & {[} -0.327, 0.122{]}  & 0.692  & {[}-0.181, 0.181{]}  & 81.606  & 1.001  & 2215.902\tabularnewline
\hline
equity / total assets  & 0.036  & {[} -0.121, 0.202{]}  & 0.692  & {[}-0.181, 0.181{]}  & 97.136  & 1.000  & 3249.143\tabularnewline
\hline
gross profit / short-term liabilities  & 0.197  & {[} -0.109, 0.501{]}  & 0.835  & {[}-0.181, 0.181{]}  & 45.970  & 1.000  & 3994.138\tabularnewline
\hline
(inventory {*} 365) / sales  & -0.197  & {[} -0.549, 0.046{]}  & 0.918  & {[}-0.181, 0.181{]}  & 50.885  & 1.001  & 2686.980\tabularnewline
\hline
operating expenses / short-term liabilities  & 9.176  & {[} 6.122, 12.508{]}  & 1.000  & {[}-0.181, 0.181{]}  & 0.000  & 1.000  & 2287.606\tabularnewline
\hline
(current assets - inventory - receivables) / short-term liabilities  & 2.720  & {[} 1.818, 3.576{]}  & 1.000  & {[}-0.181, 0.181{]}  & 0.000  & 1.000  & 2260.580\tabularnewline
\hline
profit on operating activities / sales  & -18.648  & {[}-29.352, -7.436{]}  & 1.000  & {[}-0.181, 0.181{]}  & 0.000  & 1.000  & 2439.790\tabularnewline
\hline
(current assets - inventory) / short-term liabilities  & -4.337  & {[} -5.581, -3.128{]}  & 1.000  & {[}-0.181, 0.181{]}  & 0.000  & 1.000  & 1772.371\tabularnewline
\hline
EBITDA (profit on operating activities - depreciation) / sales  & 19.113  & {[} 7.767, 29.850{]}  & 1.000  & {[}-0.181, 0.181{]}  & 0.000  & 1.000  & 2455.714\tabularnewline
\hline
long-term liabilities / equity  & -0.189  & {[} -0.562, 0.094{]}  & 0.840  & {[}-0.181, 0.181{]}  & 52.457  & 1.003  & 1579.627\tabularnewline
\hline
sales / short-term liabilities  & -10.258  & {[}-13.981, -6.939{]}  & 1.000  & {[}-0.181, 0.181{]}  & 0.000  & 1.000  & 2111.063\tabularnewline
\hline
sales / fixed assets  & 0.102  & {[} -0.011, 0.193{]}  & 0.898  & {[}-0.181, 0.181{]}  & 96.855  & 1.002  & 1982.562\tabularnewline
\hline
\end{tabular}
\end{table}
\par\end{center}

\bibliography{references1}

\begin{thebibliography}{}

\bibitem[Altman, 1968]{altman1968financial}
Altman, E.~I. (1968).
\newblock Financial ratios, discriminant analysis and the prediction of
  corporate bankruptcy.
\newblock {\em The journal of finance}, 23(4):589--609.

\bibitem[Barboza et~al., 2017]{barboza2017machine}
Barboza, F., Kimura, H., and Altman, E. (2017).
\newblock Machine learning models and bankruptcy prediction.
\newblock {\em Expert Systems with Applications}, 83:405--417.

\bibitem[Brownstein et~al., 2019]{brownstein2019role}
Brownstein, N.~C., Louis, T.~A., O'Hagan, A., and Pendergast, J. (2019).
\newblock The role of expert judgment in statistical inference and
  evidence-based decision-making.
\newblock {\em The American Statistician}, 73(sup1):56--68.

\bibitem[Carpenter et~al., 2017]{carpenter2017stan}
Carpenter, B., Gelman, A., Hoffman, M.~D., Lee, D., Goodrich, B., Betancourt,
  M., Brubaker, M., Guo, J., Li, P., and Riddell, A. (2017).
\newblock Stan: A probabilistic programming language.
\newblock {\em Journal of statistical software}, 76(1).

\bibitem[Chaudhuri and Ghosh, 2017]{chaudhuri2017bankruptcy}
Chaudhuri, A. and Ghosh, S.~K. (2017).
\newblock {\em Bankruptcy prediction through soft computing based deep learning
  technique}.
\newblock Springer.

\bibitem[Devi and Radhika, 2018]{devi2018survey}
Devi, S.~S. and Radhika, Y. (2018).
\newblock A survey on machine learning and statistical techniques in bankruptcy
  prediction.
\newblock {\em International Journal of Machine Learning and Computing},
  8(2):133--139.

\bibitem[Gelman et~al., 2013]{gelman2013bayesian}
Gelman, A., Carlin, J.~B., Stern, H.~S., Dunson, D.~B., Vehtari, A., and Rubin,
  D.~B. (2013).
\newblock {\em Bayesian data analysis}.
\newblock CRC press.

\bibitem[Hoffman and Gelman, 2014]{hoffman2014no}
Hoffman, M.~D. and Gelman, A. (2014).
\newblock The no-u-turn sampler: adaptively setting path lengths in hamiltonian
  monte carlo.
\newblock {\em J. Mach. Learn. Res.}, 15(1):1593--1623.

\bibitem[IMF, 2020]{long20206world}
IMF (2020).
\newblock World economic outlook:a long and difficult ascent.
\newblock {\em World Economic Outlook (WEO)}, Washington, DC, October.

\bibitem[Kaggle, 2020]{Companie6:online}
Kaggle (2018 (accessed October 29, 2020)).
\newblock {\em Companies bankruptcy forecast}.
\newblock \url{https://www.kaggle.com/c/companies-bankruptcy-forecast/data}.

\bibitem[Kruschke, 2014]{kruschke2014doing}
Kruschke, J. (2014).
\newblock {\em Doing Bayesian data analysis: A tutorial with R, JAGS, and
  Stan}.
\newblock Academic Press.

\bibitem[Lahmiri and Bekiros, 2019]{lahmiri2019can}
Lahmiri, S. and Bekiros, S. (2019).
\newblock Can machine learning approaches predict corporate bankruptcy?
  evidence from a qualitative experimental design.
\newblock {\em Quantitative Finance}, 19(9):1569--1577.

\bibitem[Makowski and L{\"u}decke, 2019]{makowski2019report}
Makowski, D. and L{\"u}decke, D. (2019).
\newblock The report package for r: Ensuring the use of best practices for
  results reporting.
\newblock {\em CRAN}.

\bibitem[Muth et~al., 2018]{muth2018user}
Muth, C., Oravecz, Z., and Gabry, J. (2018).
\newblock User-friendly bayesian regression modeling: A tutorial with rstanarm
  and shinystan.
\newblock {\em Quantitative Methods for Psychology}, 14(2):99--119.

\bibitem[Ohlson, 1980]{ohlson1980financial}
Ohlson, J.~A. (1980).
\newblock Financial ratios and the probabilistic prediction of bankruptcy.
\newblock {\em Journal of accounting research}, pages 109--131.

\bibitem[Snoek et~al., 2012]{snoek2012practical}
Snoek, J., Larochelle, H., and Adams, R.~P. (2012).
\newblock Practical bayesian optimization of machine learning algorithms.
\newblock In {\em Advances in neural information processing systems}, pages
  2951--2959.

\bibitem[Vehtari et~al., 2017]{vehtari2017practical}
Vehtari, A., Gelman, A., and Gabry, J. (2017).
\newblock Practical bayesian model evaluation using leave-one-out
  cross-validation and waic.
\newblock {\em Statistics and computing}, 27(5):1413--1432.

\bibitem[Vehtari et~al., 2020]{vehtari2020rank}
Vehtari, A., Gelman, A., Simpson, D., Carpenter, B., B{\"u}rkner, P.-C., et~al.
  (2020).
\newblock Rank-normalization, folding, and localization: An improved rhat for
  assessing convergence of mcmc.
\newblock {\em Bayesian Analysis}.

\bibitem[Yu et~al., 2014]{yu2014bankruptcy}
Yu, Q., Miche, Y., S{\'e}verin, E., and Lendasse, A. (2014).
\newblock Bankruptcy prediction using extreme learning machine and financial
  expertise.
\newblock {\em Neurocomputing}, 128:296--302.

\bibitem[Zi{\k{e}}ba et~al., 2016]{zikeba2016ensemble}
Zi{\k{e}}ba, M., Tomczak, S.~K., and Tomczak, J.~M. (2016).
\newblock Ensemble boosted trees with synthetic features generation in
  application to bankruptcy prediction.
\newblock {\em Expert systems with applications}, 58:93--101.

\end{thebibliography}
\bibliographystyle{apalike}


\end{document}